\documentclass[aps,preprint,amsmath,amssymb,superscriptaddress,citeautoscript,longbibliography,floatfix]{revtex4-2}

\usepackage{hyperref}
\usepackage{natbib}
\usepackage{graphicx}

\begin{document}

\title{Possible Pressure-Induced Charge-Density Wave Quantum Critical Point in LuPd$_2$In}

\author{ M.~O.~Ajeesh}
\affiliation{Max-Planck Institute for Chemical Physics of Solids, N\"{o}thnitzer Str.\ 40, 01187 Dresden}

\author{Thomas~Gruner}
\affiliation{Max-Planck Institute for Chemical Physics of Solids, N\"{o}thnitzer Str.\ 40, 01187 Dresden}
\affiliation{Cavendish Laboratory, University of Cambridge, Cambridge CB3 0HE, United Kingdom}

\author{Christoph~Geibel}
\affiliation{Max-Planck Institute for Chemical Physics of Solids, N\"{o}thnitzer Str.\ 40, 01187 Dresden}

\author{Michael~Nicklas}
\affiliation{Max-Planck Institute for Chemical Physics of Solids, N\"{o}thnitzer Str.\ 40, 01187 Dresden}

\begin{abstract}
We investigated the effect of application of hydrostatic pressure on the charge-density wave (CDW) state in Lu(Pt$_{1-x}$Pd$_x$)$_2$In by electrical-resistivity measurements. In Lu(Pt$_{0.7}$Pd$_{0.3}$)$_{2}$In we find an increase of the CDW transition temperature upon application of pressure, which is not expected based on simple volume arguments, but in line with results of a theoretical work by Kim \textit{et al.} [Phys.\ Rev.\ Lett.\ 125, 157001 (2020).]. Combining experimental and theoretical results suggests the existence of a CDW quantum critical point in stoichiometric LuPd$_2$In around $p\approx20$~GPa.
\end{abstract}

\maketitle

When a second-order phase transition is continuously suppressed to zero temperature, a quantum critical point (QCP) is formed, where quantum fluctuation may considerably affect physical properties \cite{Coleman2005,Sachdev2008}. In the vicinity of a QCP unconventional phases, such as unconventional superconductivity, might emerge eventually \cite{Mathur1998}.
So far, the majority of QCPs has been studied in magnetic materials, most of them in antiferromagnetic\cite{Mathur1998,Schroeder2000,Trovarelli2000} and only a few in ferromagnetic systems \cite{Steppke2013,Shen2020}. Only recently a QCP connected to the suppression of charge-density wave (CDW) order and the appearance of a superconducting dome has been established in the rare-earth-based Heusler system Lu(Pt$_{1-x}$Pd$_x$)$_2$In \cite{Gruner2017}. This finding was quite surprising since most CDW transitions are connected to first-order structural transitions which in many cases can also be suppressed to zero temperature but then exclude the existence of a QCP \cite{Morosan2006,Kumar2010,Zocco2015,Zhu2016}. It makes the system Lu(Pt$_{1-x}$Pd$_x$)$_2$In outstanding among CDW materials. The existence of the CDW QCP was demonstrated in Lu(Pt$_{1-x}$Pd$_x$)$_2$In by utilizing the Pd-concentration as nonthermal external control parameter.

One major disadvantage of substitution-tuning is the inevitable introduction of additional atomic disorder in the material which may strongly modify the quantum critical behavior at low temperatures. It is therefore highly desirable to find ways to study the CDW QCP in Lu(Pt$_{1-x}$Pd$_x$)$_2$In in a clean fashion.
One possible way to do that is using hydrostatic pressure as external tuning parameter. In Lu(Pt$_{1-x}$Pd$_x$)$_2$In, the CDW state appears for $x < x_c \approx 0.58$, and $T_{\rm CDW}$ rises up to $T_{\rm CDW} = 490$~K at $x = 0$ \cite{Gruner2017}.
LuPt$_2$In has a larger unit-cell volume than LuPd$_2$In which does not order down to the lowest temperatures \cite{Gruner2014,Gruner2017}. Therefore, stoichiometric LuPt$_2$In seems to be the natural choice for a pressure study of a CDW QCP in a clean system.
Recently, however, Kim \textit{et al.}\ \cite{Kim2020} reported a theoretical study on the basis of \textit{ab initio} density functional theory (DFT) calculations showing that application of hydrostatic pressure on LuPd$_2$In stabilizes the CDW state instead of suppressing it. Accordingly, they postulated a pressure induced quantum critical phase transition from a superconducting to a CDW state. This result is in contrast to the naive expectation based solely on volume arguments.
In the work of Kim \textit{et al.}\ \cite{Kim2020} the QCP is predicted to occur at $p_c \approx 28$~GPa. At such a high pressure the detection of a QCP is not easy, and providing conclusive evidence is difficult.

Here we present results of a study of the pressure dependence of the CDW transition in partially Pd substituted Lu(Pt$_{1-x}$Pd$_x$)$_2$In, which support the results of the DFT calculations by Kim \textit{et al.}\ \cite{Kim2020,Kim2018}.
For our pressure experiment we chose the $x = 0.3$ compound, since its $T_{\rm CDW}$ is already quite suppressed, with $T_{\rm CDW} = 236.5$~K at $p = 0$, while the anomaly in the resistivity $\rho(T)$ at $T_{\rm CDW}$ is still very clear and therefore provides a ideal conditions for a pressure experiment. Under pressure we find an increase of the CDW transition temperature $T_{\rm CDW}$, at a rate $dT_{\rm CDW}/dp \approx 18.3$~K/GPa. This confirms that pressure stabilizes the CDW state in the series Lu(Pt$_{1-x}$Pd$_x$)$_2$In. Combining the present data with published results on the substitution series \cite{Gruner2017}, one can estimate a $p_c$ of $19$~GPa for the onset of the CDW in LuPd$_2$In, in reasonable agreement with the predicted critical pressure $p_c \approx 28$~GPa based on DFT calculations \cite{Kim2020}.

A polycrystalline specimen of Lu(Pt$_{0.7}$Pd$_{0.3}$)$_{2}$In was synthesized by arc-melting stoichiometric amounts of the pure elements and subsequent annealing. For details of the sample preparation and characterization see Gruner \textit{et al.}\ \cite{Gruner2017}.
We studied the electrical resistivity of a sample under hydrostatic pressures up to $2.5$~GPa using a piston-cylinder-type pressure cell by a four-probe method. Silicon fluid served as pressure-transmitting medium and ensured good hydrostatic conditions at all pressures. The pressure inside the sample chamber was determined by a lead manometer. Details of the pressure cell and measurements technique can be found in Nicklas \cite{Nicklas2015}.

%
\begin{figure}[tb!]
	\begin{center}
		\includegraphics[width=0.8\columnwidth]{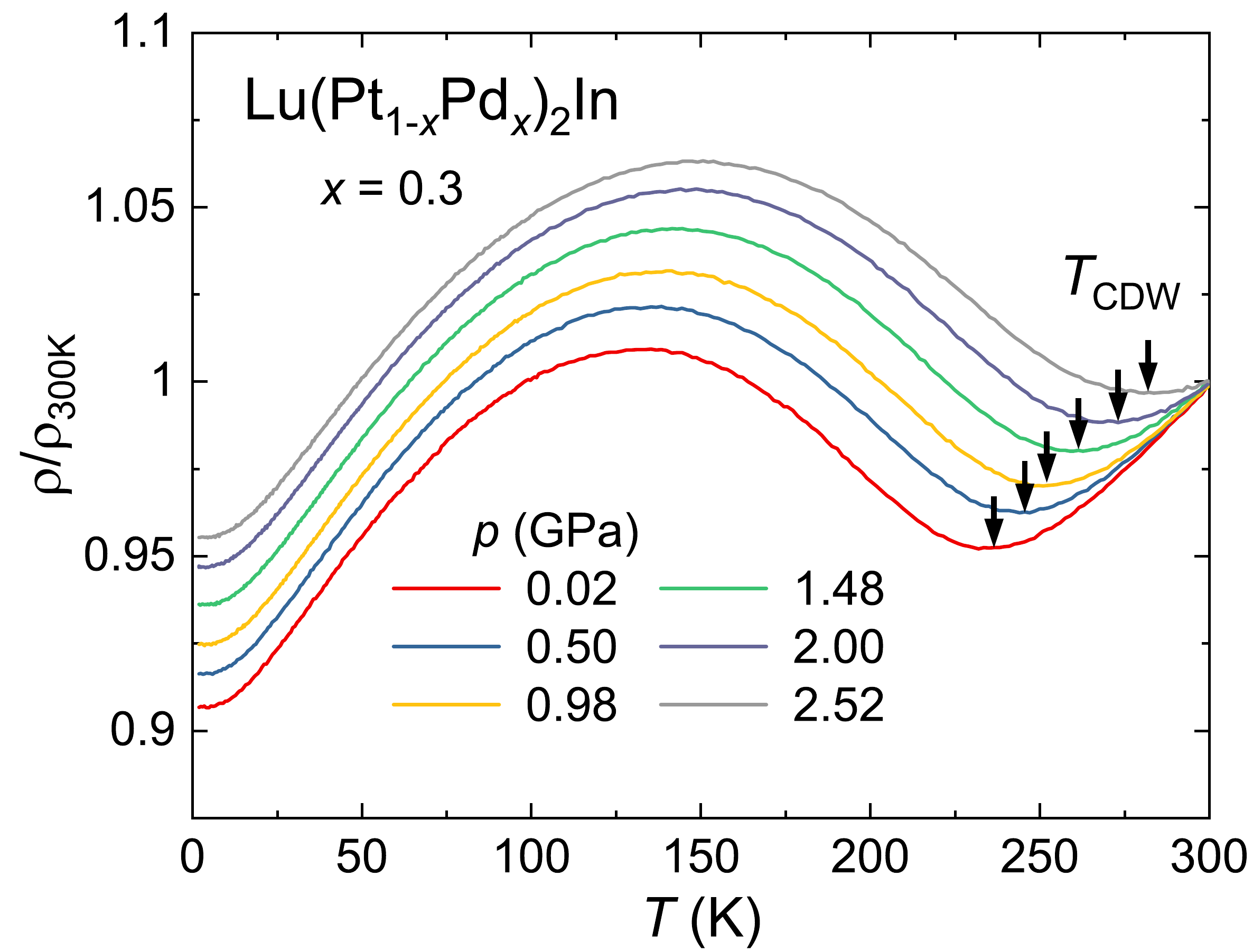}
		\caption{Resistivity of Lu(Pt$_{0.7}$Pd$_{0.3}$)$_{2}$In normalized by its value at 300~K as function of temperature for different pressures up to 2.5~GPa. Arrows mark the $T_{\rm CDW}$.
		}\label{fig1}
	\end{center}
\end{figure}

\begin{figure}[tb!]
	\begin{center}
		\includegraphics[width=0.8\columnwidth]{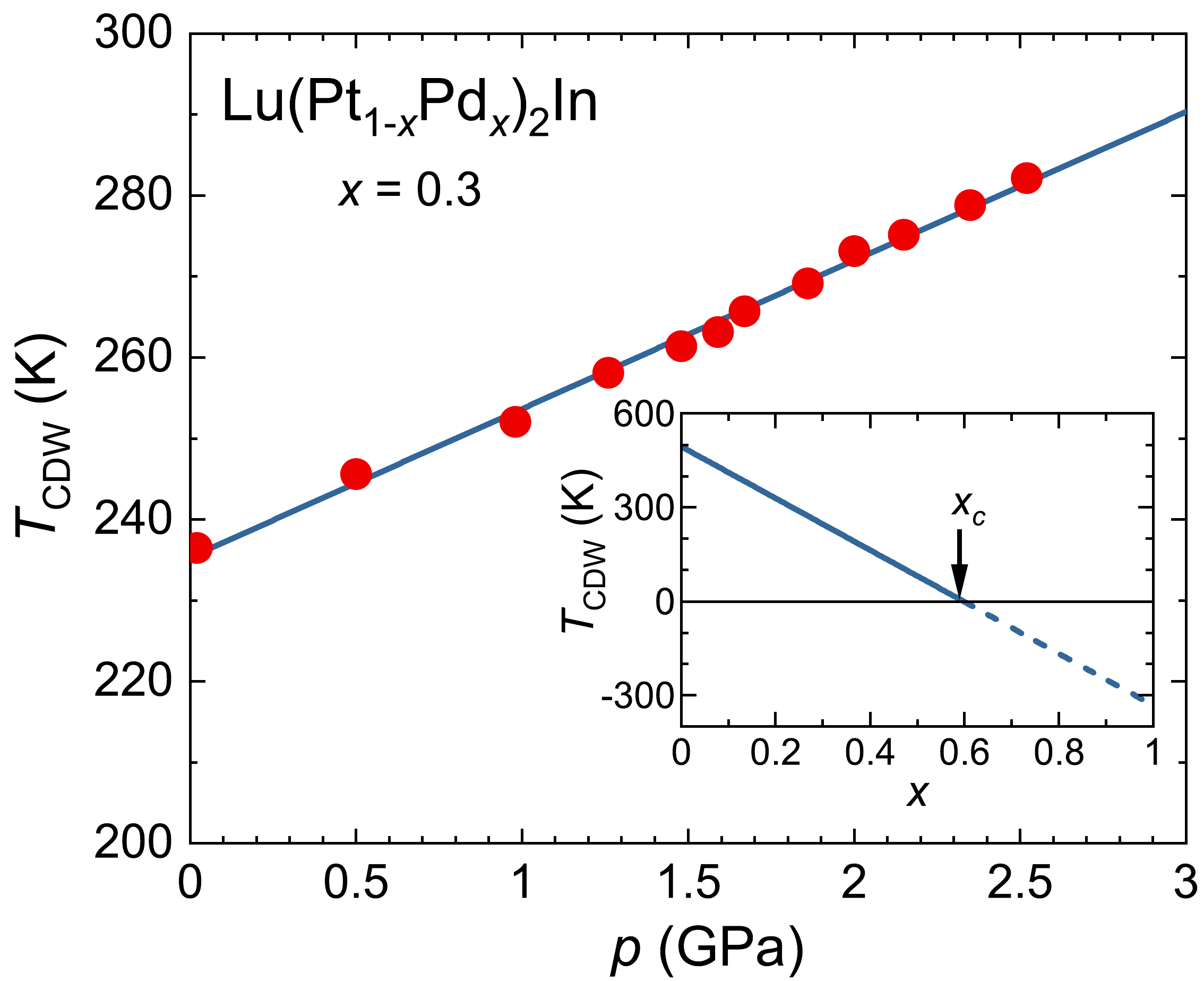}
		\caption{Temperature -- pressure phase diagram of Lu(Pt$_{0.7}$Pd$_{0.3}$)$_{2}$In. The straight line is a linear fit to the data. Inset: Temperature -- Pd-concentration phase diagram for Lu(Pt$_{1-x}$Pd$_x$)$_2$In based on the data reported by Gruner \textit{et al.}\ (solid line) \cite{Gruner2017}. The dashed line is an extrapolation to hypothetical negative $T_{\rm CDW}$ (see text for details). The arrow marks the critical concentration of the concentration-controlled CDW QCP.
		}\label{fig2}
	\end{center}
\end{figure}

In Lu(Pt$_{0.7}$Pd$_{0.3}$)$_{2}$In the opening of the CDW gap upon lowering temperature leads to an increase in resistivity (see Fig.\ \ref{fig1}. We take the minimum in $\rho(T)$ to define $T_{\rm CDW}$. Upon increasing pressure $T_{\rm CDW}(p)$ shifts linearly to higher temperatures from 236.5~K at ambient pressure to 282.2~K at 2.5~GPa (see Fig.\ \ref{fig2}). A linear fit to the data yields a slope of $dT_{\rm CDW}/dp\approx 18.3$~K/GPa. The positive pressure dependence cannot be understood by a simple volume argument. Based on the results in the Lu(Pt$_{1-x}$Pd$_x$)$_2$In substitution series, one would anticipate a suppression of $T_{\rm CDW}$ upon increasing pressure since LuPd$_2$In has a smaller unit-cell volume than LuPt$_2$In \cite{Gruner2014,Gruner2017}. The opposite is observed. That is what Kim \textit{et al.}\ found for LuPd$_{2}$In as a result of DFT calculations \cite{Kim2020}.

The DFT calculations predict that application of pressure favors the formation of a CDW in LuPd$_2$In and conclude that a CDW phase should be established above $p_c\approx28$~GPa. That is inline with the pressure-induced enhancement of $T_{\rm CDW}$  observed in Lu(Pt$_{0.7}$Pd$_{0.3}$)$_{2}$In.

Combining the experimental results of our pressure investigation with those of the substitution study by Gruner \textit{et al.}\ \cite{Gruner2017} allows for an estimation of the critical pressure for the onset of the CDW in LuPd$_2$In. In Lu(Pt$_{1-x}$Pd$_x$)$_2$In  $T_{\rm CDW}$ is linearly suppressed at a rate of $-827$~K/$x$-Pd \cite{Gruner2017}. Extrapolating this linear behavior from the critical concentration to $x = 1$ results in an hypothetical negative $T_{\rm CDW} = -337$~K in pure LuPd$_2$In, as displayed in the inset of Fig.\ \ref{fig2}. Assuming the slope $dT_{\rm CDW}/dp$ in LuPd$_2$In to be the same as that we observed in Lu(Pt$_{0.7}$Pd$_{0.3}$)$_2$In, $dT_{\rm CDW}/dp = 18.3$~K/GPa, implies a critical pressure of 19~GPa to restore the CDW in pure LuPd$_2$In. This is in reasonable agreement with the result of the DFT calculations keeping in mind the rough estimations made here.

In summary, our experimental results combined with DFT calculations reported earlier by Kim \textit{et al.}\ \cite{Kim2020} suggest the existence of a pressure-controlled CDW QCP in LuPd$_2$In in the pressure range between 15 and 30~GPa. They encourage detailed high pressure experiments on LuPd$_2$In in the relevant pressure range. These experiments are technically much more demanding than the ones presented here. To reach the estimated critical pressure diamond anvil cells have to be utilized together with dedicated low temperature equipment.

\bibliography{LuPtPdIn}

\end{document}